\documentclass[english]{revtex4-1}
\usepackage[T1]{fontenc}
\usepackage[latin9]{inputenc}
\usepackage{color}
\usepackage{amsmath}
\usepackage{amssymb}

\makeatletter

\usepackage{hyperref}
  \hypersetup{colorlinks=true,citecolor=blue}

\makeatother

\usepackage{babel}
\begin{document}

\title[NC Schrödinger-pauli equation]{{\normalsize{}The Phase-Space Noncommutativity Effect on the Large
and Small Wavefunction Components Approach at Dirac Equation}}

\author{{\normalsize{}Ilyas Houam}}

\email{ilyashaouam@live.fr}

\address{Département de Physique, Faculté des sciences exactes, Université
Des Frères Mentouri, Constantine, Algeria}
\begin{abstract}
{\normalsize{}By the large and small wave-function components approach
we achieved the nonrela-tivistic limit of the Dirac equation in interaction
with an electromagnetic potential in noncommutative phase-space, and
we tested the effect of the phase-space noncom-mutativity on it, knowing
that the nonrelativistic limit of the Dirac equation gives the Schrödinger-Pauli
equation.}{\normalsize \par}
\end{abstract}


\maketitle

\paragraph*{\textcolor{blue}{PACS:} \textcolor{blue}{03.65.Ge} Solutions of wave
equations: bound states \textendash{} 03.65.Pm Relativistic wave equations
\textendash{} 0 02.40.Gh Noncommutative geometry\textendash{} 03.65.Fd
Algebraic methods}

\section{{\normalsize{}Introduction}}

In the last few years there has been much interests in the study of
physics in non-commutative space, knowing that the study of noncommutative
geometry has a long history \cite{key-1,key-2} , the studies of noncommutativity
in phase-space and their involvement for quantum field theories play
an important role in various fields of physics especially in the theory
of strings, and in the matrix model of M-theory \cite{key-3}, also
in the description of quantum gravity. The common method for studying
the noncommutativity of quantum mechanics (NCQM) is the correspondence
between commutative space and the noncommutative space using the method
of translation which known as Bopp-shift, or using the Moyal star
product \cite{key-4,key-5,key-6,key-7}. In the quantum-mechanical
description of particles, there are various relativistic or non-relativistic
wave equations as the usual Schrödinger equation applies to the spin-0
particles in the non-relativistic domain, and the Klein\textendash Gordon
equation is the relativistic equation appropriate for spin-0 particles
\cite{key-8,key-9,key-10}, and in regards to the spin-1/2 particles
are governed by the relativistic Dirac . To move from the relativistic
quantum mechanics toward the nonrelativistic one it is very necessary
to pass by the nonrelativistic limit, which is to transform the physical
information under the condition v\ensuremath{\ll}c , we speak of the
nonrelativistic limit for low speeds in front of the speed of the
light or for the regime of weak or low energy in front of the mass
energy $\frac{p}{c}\approx\frac{pc}{mc^{2}}\ll1$ , the nonrelativistic
limit can be achieved through various ways, the most important ways
are the Foldy-Wouthuysen transformation (it is only applicable to
weak fields) \cite{key-11,key-12}, and the Douglas-Kroll-Hell transformation
\cite{key-13,key-14,key-15}, this canonical transformation is an
unitary transformation allows separating (block-diagonalize) Dirac
hamiltonian into two parts, one part describes electrons, while the
other gives rise to negative energy states, which are the so-called
positronic states, and the classical approach which is the large and
small wave function components approach \cite{key-16,key-17}, in
this work we investigate the nonrelativistic limit of the Dirac equation
according to the large and small wave function components approach
to derive directly the Schrödinger\textendash Pauli equation \cite{key-18,key-19,key-20},
but in the case of particles with spin 1 or higher, only relativistic
equations are usually considered \cite{key-21}, noting that the uses
of the Schrödinger\textendash Pauli equation represented on the study
of the fine structure of the hydrogen atom, and various scattering
problems knowing that it does not consider the spin of the particles
in the studies, but it can be introduced by assuming the presence
of an electromagnetic field in the Dirac equation befor the extraction
of the nonrelativistic equation, which describes the interaction of
a spin 1/2 particle with the external electromagnetic field. It correctly
predicts the spin of the particle and the gyromagnetic ratio, in fact
the examples of using and applying Schrödinger\textendash Pauli equation
are many and we can not all mention them.

\section{{\normalsize{}length-momentum noncommutativity}}

At string scales (very small scales) the space does not commute anymore,
so that we consider the operators of coordinates and momentum in the
noncommutative phase-space $x_{i}^{nc}$ and $p_{i}^{nc}$ respectivly,
then considering a noncommutative algebra satisfying the commutation
relations 
\begin{equation}
\begin{array}{c}
\left[x_{i}^{nc},x_{j}^{nc}\right]=i\Theta_{ij}\,\,\,\\
\left[p_{i}^{nc},p_{j}^{nc}\right]=i\overline{\Theta}_{ij}\,\,\,\\
\left[x_{i}^{nc},p_{i}^{nc}\right]=i\hbar^{eff}\delta_{ij}
\end{array},\,\:(i,j=1,2),\label{eq:2-1}
\end{equation}

taking into account the effective Plank constant
\begin{equation}
\hbar^{eff}=\hbar(1+\frac{\Theta\bar{\Theta}}{4\hbar^{2}}).\label{eq:3-1}
\end{equation}

Where $\Theta_{ij}=\epsilon_{ijk}\Theta_{k}\,,\:\Theta_{k}=(0,0,\Theta)\,,\,\overline{\Theta}_{ij}=\epsilon_{ijk}\overline{\Theta}_{k}\,,\:\overline{\Theta}_{k}=(0,0,\overline{\Theta})$,
$\Theta\,,\:\overline{\Theta}$ are antisymmetric constant matrices
(noncommutative parameters ) with the dimension of $(lenght)^{2}$and
$(momentum)^{2}$, respectively.

The mapping between the noncommutative phase space and the commutative
one doing through the \textit{Bopp-shift} linear transformations \cite{key-22,key-23}

\begin{equation}
\begin{array}{c}
x^{nc}=x-\frac{1}{2\hbar}\Theta p_{y}\\
y^{nc}=y+\frac{1}{2\hbar}\Theta p_{x}\\
p_{x}^{nc}=p_{x}+\frac{1}{2\hbar}\bar{\Theta}y\\
p_{y}^{nc}=p_{y}-\frac{1}{2\hbar}\bar{\Theta}x
\end{array},\label{eq:4-1}
\end{equation}

vanishing the noncommutative parameters the system will reduce to
the commutative one.

With another method the noncommutativity in space can be realized
using the \textit{Moyal product} ( $\star-product$ ), \cite{key-24,key-25,key-26} 

\begin{equation}
\begin{array}{c}
(f\star g)(x)=\exp[\frac{i}{2}\theta_{ab}\partial_{x_{a}}\partial_{x_{b}}]f\left(x_{a}\right)g\left(x_{b}\right)=f(x)g(x)\\
+\sum_{n=1}\left(\frac{1}{n!}\right)\left(\frac{i}{2}\right)^{n}\theta^{a_{1}b_{1}}...\theta^{a_{n}b_{n}}\partial_{a_{1}}...\partial_{a_{k}}f(x)\partial_{b_{1}}...\partial_{b_{k}}g(x),
\end{array}\label{eq:5-1}
\end{equation}

in other term the noncommutative information is encoded in the star
product

\begin{equation}
\left(A,\star\right)\cong\left(A^{nc},\cdot\right)\label{eq:6-1}
\end{equation}

\section{{\normalsize{}3. Nonrelativistic Limit of the Noncommutative Dirac
Equation}}

\subsection{{\normalsize{}Noncommutative Dirac Equation}}

Starting with Dirac equation in the noncommutative phase-space \cite{key-27,key-28}
\begin{equation}
H(x,p)\star\psi(x)=H(x^{nc},p^{nc}).\psi(x^{nc})=E\psi,\label{eq:19}
\end{equation}

the Dirac equation in interaction with electromagnetic four-potential
\{$A_{\mu}=\left(A_{0}(x),\:A_{i}(x)\right)$\} in commutative space-phase
is 
\begin{equation}
\left\{ c\alpha_{i}(\hat{p_{i}}-\frac{e}{c}A_{i}(x))+eA_{0}(x)+\hat{\beta}mc^{2}\right\} \psi=E\psi,\label{eq:7-1}
\end{equation}

where the momentum $\hat{p_{i}}$ is given by $\hat{p_{i}}=-i\hbar\nabla_{i}$
and the matrices $\alpha{}_{i}$ and $\beta$ satisfy the anticommutation
relations
\begin{equation}
\left\{ \alpha_{i},\alpha_{j}\right\} =2\delta_{ij}\:,\;\left\{ \alpha_{i},\beta\right\} =0\:,\;\alpha_{i}^{2}=\beta^{2}=1.\label{eq:7-2}
\end{equation}

Using the Eq.(\ref{eq:5-1}), we achieve the noncommutativity in space
$\left\{ x\rightarrow x^{nc}\right\} $, then the Dirac equation Eq.(\ref{eq:7-1})
becomes

\begin{equation}
\left\{ c\alpha_{i}(\hat{p_{i}}-\frac{e}{c}A_{i}(x))+eA_{0}(x)+\hat{\beta}mc^{2}\right\} \star\psi(x)=E\psi_{nc},\label{eq:20-2}
\end{equation}

as $A(x)=hx$ for that the derivation in the Eq.(\ref{eq:5-1}) will
automatically stop in the first order, 

\begin{equation}
(f\star g)(x)=f(x)g(x)+\frac{i}{2}\Theta^{ab}\partial_{a}f\partial_{b}g+\mathcal{O}(\theta^{2}),\label{eq:20-3}
\end{equation}

it means that terms upper than the first order will vanish, then the
Eq.(\ref{eq:20-2}) transforms to

\begin{equation}
(H\star\psi)(x)=H(x,p^{nc})\psi(x)+\frac{i}{2}\Theta_{ab}\partial_{a}\left(c\alpha_{i}(\hat{p_{i}}-\frac{e}{c}A_{i}(x))+eA_{0}(x)+\hat{\beta}mc^{2}\right)\partial_{b}\psi(x)=E\psi_{nc},\label{eq:20}
\end{equation}

as $\partial_{a}(c\alpha_{i}\hat{p_{i}})=\partial_{a}(\hat{\beta}mc^{2})=0$
Eq.(\ref{eq:20}) becomes
\begin{equation}
H(x,p^{nc)})\psi(x)-\frac{ie}{2}\Theta_{ab}\partial_{a}\left(\alpha_{i}A_{i}(x))+A_{0}(x)\right)\partial_{b}\psi(x)=E\psi(x).\label{eq:21}
\end{equation}

after there we achieve the noncommutativity in phase $\left\{ p\rightarrow p^{nc}\right\} $using
the Eq.(\ref{eq:4-1}) for finding the entire noncommutative phase-space
Dirac equation,
\begin{equation}
\begin{array}{c}
H_{nc}(x,p)\star\psi(x)=\left\{ c\alpha_{i}\left(p_{i}+\frac{1}{2\hbar}\overline{\Theta}_{ij}x_{j}-\frac{e}{c}A_{i}(x)\right)+eA_{0}(x)\right.\\
\left.+\hat{\beta}mc^{2}-\frac{ie}{2}\Theta_{ab}\partial_{a}\left(\alpha_{i}A_{i}(x)+A_{0}(x)\right)\partial_{b}\right\} \psi(x)=E\psi_{nc},
\end{array}\label{eq:22}
\end{equation}

we rewrite Eq.(\ref{eq:22}) in a more compact form (see \textbf{\textit{Appendix}}
):
\begin{equation}
\begin{array}{c}
H_{nc}\star\psi_{nc}=\left[c\overrightarrow{\alpha}\left(\hat{\overrightarrow{p}}-\frac{e}{c}\overrightarrow{A}\right)+eA_{0}+\hat{\beta}mc^{2}\right.\\
\left.+\frac{c}{\hbar}(\overrightarrow{\alpha}\times\overrightarrow{x}).\overrightarrow{\overline{\Theta}}+\frac{e}{\hbar}\left(\overrightarrow{\nabla}\left(\overrightarrow{\alpha}\overrightarrow{A}-A_{0}\right)\times\overrightarrow{p}\right).\overrightarrow{\Theta}\right]\psi_{nc}=E\psi_{nc}.
\end{array}\label{eq:23}
\end{equation}

\subsection{{\normalsize{}Large and Small Wave-function Components Approach}}

It is possible to define the nonrelativistic limit of the Dirac equation,
using several ways, including that there is the Douglas-Kroll-Hell
approach, it used mostly as part of relativistic quantum chemistry
,and the Foldy-Wouthuysen transformation, which are both canonical
transformation, and the method of development in power of $\hbar$
\cite{key-29}, and the classical approach, the latter one depends
on the upper two components of the Dirac wave-function $\psi$ in
the standard representation are much larger than the lower two components,
using this property we can derive simply the Schrödinger-Pauli equation.

To define the nonrelativistic limit of the phase-space noncommutative
Dirac equation we should firstly study the case of an electron at
rest, so that without the electromagnetic interaction,\{$\hat{\overrightarrow{p}}\psi=0,\:A^{\mu}=0$\}
Eq.(\ref{eq:23}) becomes
\begin{equation}
H_{nc}\psi_{nc}=\left\{ \hat{\beta}m_{0}c^{2}+\frac{c}{\hbar}(\overrightarrow{\alpha}\times\overrightarrow{x}).\overrightarrow{\overline{\Theta}}\right\} \psi_{nc}=i\hbar\frac{\partial\psi_{nc}}{\partial t}.\label{eq:24}
\end{equation}

This system of equations is simply solved, and leads to the following
four-solutions
\begin{equation}
\begin{array}{cc}
\psi_{nc}^{1}=\left(\begin{array}{c}
1\\
0\\
0\\
0
\end{array}\right)e^{-\frac{i}{\hbar}(m_{0}c^{2}+\frac{c}{\hbar}(\overrightarrow{\alpha}\times\overrightarrow{x}).\overrightarrow{\overline{\Theta}})t} & \psi_{nc}^{2}=\left(\begin{array}{c}
0\\
1\\
0\\
0
\end{array}\right)e^{-\frac{i}{\hbar}(m_{0}c^{2}+\frac{c}{\hbar}(\overrightarrow{\alpha}\times\overrightarrow{x}).\overrightarrow{\overline{\Theta}})t}\\
\psi_{nc}^{3}=\left(\begin{array}{c}
0\\
0\\
1\\
0
\end{array}\right)e^{+\frac{i}{\hbar}(m_{0}c^{2}+\frac{c}{\hbar}(\overrightarrow{\alpha}\times\overrightarrow{x}).\overrightarrow{\overline{\Theta}})t} & \psi_{nc}^{4}=\left(\begin{array}{c}
0\\
0\\
0\\
1
\end{array}\right)e^{+\frac{i}{\hbar}(m_{0}c^{2}+\frac{c}{\hbar}(\overrightarrow{\alpha}\times\overrightarrow{x}).\overrightarrow{\overline{\Theta}})t},
\end{array}\label{eq:25}
\end{equation}

$\psi_{nc}^{1}$and $\psi_{nc}^{2}$correspond to the positive energy
value and $\psi_{nc}^{3}$, $\psi_{nc}^{4}$to the negative one.

At first therefore we restrict ourselves to solutions of positive
energy. In order to show that the Dirac equation reproduces the two
component Pauli equation in the nonrelativistic limit.

The nonrelativistic limit of the Eq.(\ref{eq:23}) can be most efficiently
studied in the representation
\begin{equation}
\psi_{nc}=\left(\begin{array}{c}
\tilde{\varphi}_{nc}\\
\tilde{\chi}_{nc}
\end{array}\right),\label{eq:25-1}
\end{equation}

where the four-component spinor $\psi_{nc}$ is decomposed into two-two
component spinors $\tilde{\varphi}$and $\tilde{\chi}$, with \{$\hat{\overrightarrow{p}}-\frac{e}{c}\overrightarrow{A}\rightarrow\hat{\overrightarrow{\Pi}}$\},
the Dirac equation Eq.(\ref{eq:23}) becomes 
\begin{equation}
\begin{array}{c}
i\hbar\frac{\partial}{\partial t}\left(\begin{array}{c}
\tilde{\varphi}_{nc}\\
\tilde{\chi}_{nc}
\end{array}\right)=c\overrightarrow{\alpha}\hat{\overrightarrow{\Pi}}\left(\begin{array}{c}
\tilde{\varphi}_{nc}\\
\tilde{\chi}_{nc}
\end{array}\right)+eA_{0}\left(\begin{array}{c}
\tilde{\varphi}_{nc}\\
\tilde{\chi}_{nc}
\end{array}\right)+\hat{\beta}m_{0}c^{2}\left(\begin{array}{c}
\tilde{\varphi}_{nc}\\
\tilde{\chi}_{nc}
\end{array}\right)\\
+\frac{c}{\hbar}(\overrightarrow{\alpha}\times\overrightarrow{x}).\overrightarrow{\overline{\Theta}}\left(\begin{array}{c}
\tilde{\varphi}_{nc}\\
\tilde{\chi}_{nc}
\end{array}\right)+\frac{e}{\hbar}\left(\overrightarrow{\nabla}\left(\overrightarrow{\alpha}\overrightarrow{A}-A_{0}\right)\times\overrightarrow{p}\right).\overrightarrow{\Theta}\left(\begin{array}{c}
\tilde{\varphi}_{nc}\\
\tilde{\chi}_{nc}
\end{array}\right),
\end{array}\label{eq:26}
\end{equation}

according to the Dirac matrices 
\begin{equation}
\hat{\overrightarrow{\alpha}}=\left(\begin{array}{cc}
0 & \hat{\overrightarrow{\sigma}}\\
\hat{\overrightarrow{\sigma}} & 0
\end{array}\right)\beta=\left(\begin{array}{cc}
1 & 0\\
0 & -1
\end{array}\right),\label{eq:27-1}
\end{equation}

and setting $\varTheta_{\overline{\Theta}}=(\overrightarrow{\alpha}\times\overrightarrow{x}).\overrightarrow{\overline{\Theta}}\:$and
$\varTheta_{\Theta}=(\overrightarrow{\nabla}(\overrightarrow{\alpha}\overrightarrow{A}-A_{0})\times\overrightarrow{p}).\overrightarrow{\Theta}$
it comes
\begin{equation}
\begin{array}{c}
i\hbar\frac{\partial}{\partial t}\left(\begin{array}{c}
\tilde{\varphi}_{nc}\\
\tilde{\chi}_{n}
\end{array}\right)=\left(\begin{array}{c}
c\hat{\overrightarrow{\sigma}}\hat{\overrightarrow{\Pi}}\tilde{\chi}_{nc}\\
c\hat{\overrightarrow{\sigma}}\hat{\overrightarrow{\Pi}}\tilde{\varphi}_{nc}
\end{array}\right)+eA_{0}\left(\begin{array}{c}
\tilde{\varphi}_{nc}\\
\tilde{\chi}_{nc}
\end{array}\right)+m_{0}c^{2}\left(\begin{array}{c}
\tilde{\varphi}_{nc)}\\
-\tilde{\chi}_{nc}
\end{array}\right)\\
+\frac{c}{\hbar}\varTheta_{\overline{\Theta}}\left(\begin{array}{c}
\tilde{\varphi}_{nc}\\
\tilde{\chi}_{nc)}
\end{array}\right)+\frac{e}{\hbar}\varTheta_{\Theta}\left(\begin{array}{c}
\tilde{\varphi}_{nc}\\
\tilde{\chi}_{nc}
\end{array}\right),
\end{array}\label{eq:27}
\end{equation}

if the rest energy $m_{0}c^{2},$as the largest occuring energy, is
additionally separated by$\left(\begin{array}{c}
\tilde{\varphi}_{nc}\\
\tilde{\chi}_{nc}
\end{array}\right)=\left(\begin{array}{c}
\varphi_{nc}\\
\chi_{nc}
\end{array}\right)e^{-\frac{i}{\hbar}(m_{0}c^{2}+\frac{c}{\hbar}\varTheta_{\overline{\Theta}})t}$ then Eq.(\ref{eq:27}) takes the form 
\begin{equation}
i\hbar\frac{\partial}{\partial t}\left(\begin{array}{c}
\varphi_{nc}\\
\chi_{nc}
\end{array}\right)=\left(\begin{array}{c}
c\hat{\overrightarrow{\sigma}}\hat{\overrightarrow{\Pi}}\chi_{nc}\\
c\hat{\overrightarrow{\sigma}}\hat{\overrightarrow{\Pi}}\varphi_{nc}
\end{array}\right)+eA_{0}\left(\begin{array}{c}
\varphi_{nc}\\
\chi_{nc}
\end{array}\right)-2m_{0}c^{2}\left(\begin{array}{c}
0\\
\chi_{nc}
\end{array}\right)+\frac{e}{\hbar}\varTheta_{\Theta}\left(\begin{array}{c}
\varphi_{nc}\\
\chi_{nc}
\end{array}\right).\label{eq:28}
\end{equation}

firstly considering the lower of the above equation. Using the slow-time
dependence $E_{0}\gg i\hbar\frac{\partial}{\partial t}$ , and the
weak coupling of the electromagnetic potential $E_{0}\gg eA_{0}$
approach, which means that the kinetic energy as well as the potential
energy are small compared to the rest energy, by another term the
transition to the nonrelativistic limit is realized by assuming that
the momentum is small compared to the characteristic quantity mc and
that the Coulomb interaction energy is weak compared to the mass energy,
so that the Eq.(\ref{eq:28}) goes to

\begin{equation}
\left(\begin{array}{c}
c\hat{\overrightarrow{\sigma}}\hat{\overrightarrow{\Pi}}\chi_{nc}\\
c\hat{\overrightarrow{\sigma}}\hat{\overrightarrow{\Pi}}\varphi_{nc}
\end{array}\right)-2m_{0}c^{2}\left(\begin{array}{c}
0\\
\chi_{nc}
\end{array}\right)+\frac{e}{\hbar}\varTheta_{\Theta}\left(\begin{array}{c}
\varphi_{nc}\\
\chi_{nc}
\end{array}\right)=0.\label{eq:29}
\end{equation}

Using the second equation of the above system Eq.(\ref{eq:29}) then
we obtain
\begin{equation}
\chi_{nc}=\frac{c\hat{\overrightarrow{\sigma}}\hat{\overrightarrow{\Pi}}}{2m_{0}c^{2}-\frac{e}{\hbar}\varTheta_{\Theta}}\varphi_{nc}\label{eq:30}
\end{equation}
where $\chi_{nc}$ represent the small component of the wave function
$\psi_{nc}$ . Insertion of Eq.(\ref{eq:30}) into the first equation
of Eq.(\ref{eq:28}) results in a nonrelativistic wave function for
$\varphi_{nc}$
\begin{equation}
i\hbar\frac{\partial}{\partial t}\varphi_{nc}=\frac{(\hat{\overrightarrow{\sigma}}\hat{\overrightarrow{\Pi}})(\hat{\overrightarrow{\sigma}}\hat{\overrightarrow{\Pi}})}{2m_{0}-\frac{e}{\hbar c^{2}}\varTheta_{\Theta}}\varphi_{nc}+eA_{0}\varphi_{nc}+\frac{e}{\hbar}\varTheta_{\Theta}\varphi_{nc}.\label{eq:31}
\end{equation}

with the help of 
\begin{equation}
(\hat{\overrightarrow{\sigma}}\hat{\overrightarrow{A}})(\hat{\overrightarrow{\sigma}}\hat{\overrightarrow{B}})=\hat{\overrightarrow{A}}.\hat{\overrightarrow{B}}+i\hat{\overrightarrow{\sigma}}.(\hat{\overrightarrow{A}}\times\hat{\overrightarrow{B}}).\label{eq:31-1}
\end{equation}

Finally the Eq.(\ref{eq:31}) becomes
\begin{equation}
i\hbar\frac{\partial}{\partial t}\varphi_{(NC)}=\left[\frac{(\hat{\overrightarrow{p}}-\frac{e}{c}\overrightarrow{A})^{2}}{2m_{0}-\frac{e}{\hbar c^{2}}\varTheta_{\Theta}}-\frac{e\hbar\hat{\overrightarrow{\sigma}}.\overrightarrow{B}}{c\left(2m_{0}-\frac{e}{\hbar c^{2}}\varTheta_{\Theta}\right)}+eA_{0}+\frac{e}{\hbar}\varTheta_{\Theta}\right]\varphi_{(NC)}.\label{eq:31-2}
\end{equation}

This is as it should be, \textit{the noncommutative phase-space Schrödinger-Pauli
equation}, 

for $\theta=0$$\Rightarrow\varTheta_{\Theta}=0$ , the Eq.(\ref{eq:31-2})
returns to that of usual Schrödinger-Pauli equation\cite{key-30,key-31}

\subsection{{\normalsize{}Gyromagnetic Factor of The Electron (}\textit{\normalsize{}g=2}{\normalsize{})}}

According to $\overrightarrow{B}=\overrightarrow{\nabla}\times\overrightarrow{A}$,
$\overrightarrow{A}=\frac{1}{2}\overrightarrow{B}\times\overrightarrow{x}$
we have
\begin{equation}
(\hat{\overrightarrow{p}}-\frac{e}{c}\overrightarrow{A})^{2}=(\hat{\overrightarrow{p}}-\frac{e}{2c}\overrightarrow{B}\times\overrightarrow{x})^{2}\approx\hat{\overrightarrow{p}}^{2}-\frac{e}{c}\overrightarrow{B}.\overrightarrow{L},\label{eq:17}
\end{equation}

where$\overrightarrow{L}=\overrightarrow{x}\times\hat{\overrightarrow{p}}$
and $\hat{\overrightarrow{S}}=\frac{1}{2}\hbar\hat{\overrightarrow{\sigma}}$
are the operator of orbital angular momentum and the spin operator
respectively.

So that Eq.(\ref{eq:31-2}) finally takes the form of
\begin{equation}
i\hbar\frac{\partial}{\partial t}\varphi_{nc}=\left[\frac{\hat{\overrightarrow{p}}^{2}}{2m_{0}-\frac{e}{\hbar c^{2}}\varTheta_{\Theta}}-\frac{e}{c\left(2m_{0}-\frac{e}{\hbar c^{2}}\varTheta_{\Theta}\right)}(\overrightarrow{L}+2\hat{\overrightarrow{S}}).\overrightarrow{B}+eA_{0}+\frac{e}{\hbar}\varTheta_{\Theta}\right]\varphi_{nc}.\label{eq:32}
\end{equation}

while we are in very tiny space scales, so the Nc term$\varTheta_{\Theta}\ll1$
, it is possible to use the Maclaurin series, by changing the variable
\begin{equation}
\frac{e}{2m_{0}\hbar c^{2}}\varTheta_{\Theta}=\mathbf{\mathbf{\varOmega_{\Theta}}}\label{eq:33-1}
\end{equation}
\begin{equation}
\frac{1}{2m_{0}\left(1-\frac{e}{2m_{0}\hbar c^{2}}\varTheta_{\Theta}\right)}\backsimeq\frac{1}{2m_{0}}\overset{n}{\underset{j=0}{\sum}}\varOmega_{\Theta}^{j},\label{eq:33}
\end{equation}

we find that Eq.(\ref{eq:32}) goes to
\begin{equation}
i\hbar\frac{\partial}{\partial t}\varphi_{nc}=\left[\frac{1}{2m_{0}}\overset{n}{\underset{j=0}{\sum}}(e\varOmega_{\Theta}^{j}.\hat{\overrightarrow{p}}^{2}-\varOmega_{\Theta}^{j}.\frac{(\overrightarrow{L}+2\hat{\overrightarrow{S}}).\overrightarrow{B}}{c})+eA_{0}+2m_{0}c^{2}\Theta_{\theta}\right]\varphi_{nc}.\label{eq:34}
\end{equation}

The Eq.(\ref{eq:32}) represents \textit{The phase-space noncommutative
Schrödinger-Pauli equation, }and it contains the NC kinetic energy
operator and the NC Zeeman coupling term (which had been added by
hand by Pauli when we talk about the commutative term), and the term
that associated with the NC diamagnetism, in the absence of magnetic
field $(A=B=0)$, the Eq.(\ref{eq:34}) takes its original form without
the information about the spin, which is the noncommutative Schrödinger
equation as follows 

\begin{equation}
i\hbar\frac{\partial}{\partial t}\varphi_{nc}=\left[\frac{1}{2m_{0}}\overset{n}{\underset{j=0}{\sum}}e\varOmega_{\Theta}^{j}.\hat{\overrightarrow{p}}^{2}+2m_{0}c^{2}\Theta_{\theta}\right]\varphi_{nc}.\label{eq:34-1}
\end{equation}

The Eq.(\ref{eq:34}) is a first order equation of 1/m, the nonrelativistic
expansion of this equation allows to add potentials such the electrical
potential, but also to find corrections terms if one realize the development
in the second and third order of 1/m, precisely we predict that in
the seconde order we find the Darwin interaction and Spin-Orbit coupling
NC terms knowing that Darwin's NC term is interpreted as a correction
of the potential energy due to the Zitterbewegung phenomenon (the
tremor movement) \cite{key-32,key-33}, in the third order we find
corrections of the kinetic energy and temporal dependence of the electric
field NC terms.

\section{{\normalsize{}Conclusion}}

In conclusion the nonrelativistic limit of the Dirac equation with
electromagnetic potential has been studied in noncommutative phase-space
using the large and small wavefunction components approach. We find
that the effect of the noncommutativity in phase on the nonrelativistic
limit vanished, but the effect of the noncommutativity in space appeared
widely and it reduced in the $\varTheta_{\Theta}$ (at least to the
order of approximation we have considered). Under the condition that
space-space and momentum-momentum are all commutative (namely, $\overline{\Theta}$=
0, $\Theta$= 0) the results return to that of the usual quantum mechanics. 
\begin{acknowledgments}
The author would like to thank Pr Lyazid Chetouani for interesting
comments and suggestions.
\end{acknowledgments}

\appendix

\section*{{\normalsize{}Appendix A.}}

calculations between moving from the relation Eq.(\ref{eq:22}) to
the relation Eq.(\ref{eq:23})

$\vphantom{}$

using $\eta_{ij}=\eta\epsilon_{ij}$ and $\eta_{k}=\frac{1}{2}\epsilon_{kij}\eta_{ij}$

$\vphantom{}$

$c\alpha_{i}\frac{1}{2\hbar}\eta_{ij}x_{j}=c2\frac{1}{2\hbar}\eta_{k}\epsilon_{kij}^{-1}\alpha_{i}x_{j}=c\frac{1}{\hbar}\eta_{k}\epsilon_{kij}\alpha_{i}x_{j}$
/ $\epsilon_{kij}=\epsilon_{ijk}$

$\vphantom{}$

we know that $\left(u\times v\right)_{\mu}=\epsilon_{\mu\nu\lambda}u_{\nu}.v_{\lambda}$so

$\vphantom{}$

$c\frac{1}{\hbar}\eta_{k}\epsilon_{kij}\alpha_{i}x_{j}=\frac{c}{\hbar}\left(\overrightarrow{\alpha}\times\overrightarrow{x}\right)_{k}.\eta_{k}=\frac{c}{\hbar}\left(\overrightarrow{\alpha}\times\overrightarrow{x}\right).\overrightarrow{\eta}$

$\vphantom{}$

with the same manner we prove that

$\vphantom{}$

$-ie\theta_{k}\epsilon_{abk}\partial_{a}\left(\overrightarrow{\alpha}\overrightarrow{A}-A_{0}\right)\partial_{b}=-ie\frac{\hbar}{\hbar}\theta_{k}\epsilon_{abk}\partial_{a}\left(\overrightarrow{\alpha}\overrightarrow{A}-A_{0}\right)=\frac{e}{\hbar}\left(\overrightarrow{\nabla}.\left(\overrightarrow{\alpha}\overrightarrow{A}-A_{0}\right)\times\overrightarrow{p}\right).\overrightarrow{\theta}$

\end{document}